\documentclass[11pt,eqs]{article}
\usepackage{latexsym}
\usepackage{amsmath}
\usepackage{hyperref}
\hypersetup{
	colorlinks=true,
	linkcolor=cyan,
	filecolor=magenta,      
	urlcolor=cyan,
}

\makeatletter
\newcommand\xleftrightarrow[2][]{%
	\ext@arrow 9999{\longleftrightarrowfill@}{#1}{#2}}
\newcommand\longleftrightarrowfill@{%
	\arrowfill@\leftarrow\relbar\rightarrow}
\makeatother

\title{
	T-duality of a bosonic string in a weakly curved space-time 
	\thanks{Work supported in part by the Serbian Ministry of Education,Science and Technological Development
		}}
\author{S. Dedi\'c, D. Obri\'c \thanks{email: sonjadedic@protonmail.com, dobric@ipb.ac.rs}\\ {\it Institute of Physics, University of Kreagujevac, Radoja Domanovi\'ca 12, Serbia} \\ {\it Institute of Physics Belgrade, University of Belgrade, Pregrevica 118, Serbia} }

\begin{document}
	\maketitle
	\begin{abstract}
		In this article we consider T-dualization of a $3D$ closed bosonic string that is propagating in space-time metric that has infinitesimal linear dependence on the coordinates $x^\mu$. Other fields, Kalb-Ramond and dilaton fields are set to zero. Action with this configuration of fields is not invariant to translations. In order to find T-dual theory we will employ generalization of the Buscher procedure that can be applied to such cases where we have coordinate dependent fields that do not possess translational isometry. Finally, using transformation laws that connect coordinates of starting and T-dual theories, we will be able to examine the geometric structure of T-dual theory.
	\end{abstract}

\section{Introduction}
\setcounter{equation}{0}

String theory is currently one of the best possible candidates for the theory of everything, theory that has the capacity to explain all possible physical phenomena. The framework of string theory is such that it offers a description of both gauge and gravity interactions, making it possible to examine models which have been impossible for standard quantum mechanics \cite{Zwiebach,Polchinski,Polchinski2,Becker Becker Schwarz,book2}. In addition to describing all fundamental interactions, string theory possesses some additional peculiar properties, some of which are presence of two dualities, T and S duality, that connect seemingly different models of string theory. First of these dualities, T-duality, connects theories with radii of compactified dimension $R$ with ones where radii of compactified dimension is $1/R$ \cite{T-duality procedure 1,Rocek Verlinde,T-duality explained 2,Alvarez Alvarez-Gaume Barbon Lozano}. Second duality, S-duality, connects theories that have strength of coupling constant $g$ with theories that have strength of coupling constant $1/g$. These two dualities imply existence of a theory more fundamental that string theory, M theory, where different types of string theories emerge as a limiting cases of this more fundamental theory \cite{Polchinski2,Becker Becker Schwarz,book2}.

One of possible ways to obtain T-dual theories is by Buscher procedure \cite{T-duality procedure 1,T-duality procedure 2}. This procedure relies on existence of underlying global isometries in the given model. Basically, in order to find T-dual theory for a given theory, we need to localize some global isometry, usually translational symmetry. This is accomplished by replacing all derivatives with covariant ones and in cases where we have coordinate dependent background fields we also replace all coordinates with invariant ones \cite{Lj. B. S. T-dualizacija 1,Generalized Buscher procedure,Davidovic Nikolic Sazdovic,ljdbs}. Covariant derivatives introduce additional degrees of freedom as gauge fields. Since both starting and dual theory have to describe same phenomena, we require that both theories have the same number of degrees of freedom. By introducing Lagrange multipliers into the action we can eliminate newly created degrees of freedom. Utilizing gauge freedom we can fix starting coordinates, thus obtaining action that is only a function of gauge fields and Lagrange multipliers. Finding equations of motion for Lagrange multipliers and inserting their solutions into gauge fixed action we are able to return to the starting theory. Finding equations of motion for gauge fields and inserting their solutions into gauge fixed action we obtain T-dual theory. While Buscher procedure works for theories that posses global isometries, it is possible to extend this procedure to cases where such isometries do not exist \cite{Auxiliary action Buscher procedure}. This extension works by replacing starting action with one that has translational symmetry. This "replacement" action is the same as the starting one where we have replaced derivatives with covariant ones, coordinates with invariant ones, introduced Lagrange multipliers and where we have fixed gauge freedom. Validity of this move is justified if we are able to obtain starting action by inserting solutions to equations of motion for Lagrange multipliers in replacement action. 

One of the useful features of T-duality is the ability to examine what happens to underlying geometry of space-time in dual theory if we have specific configuration of background fields in starting theory. For example, it has been shown that closed bosonic strings in presence of coordinate dependent Kalb-Ramond field have non-commutative dual theory \cite{Nikolic Obric}, where depending on the order in which T-duality has been performed there is an emergence of a whole class of non-commutative and non-associative theories \cite{Integration of Poisson brackets 2}. While non-commutativity in the context of open string theory is nothing new \cite{Open string noncommutativity 1,sazda1,sazda3,Open string noncommutativity 2,Open string noncommutativity 3,Open string noncommutativity 4,Open string noncommutativity 5,Open string noncommutativity 6,Open string noncommutativity 7,Open string noncommutativity 8,Open string noncommutativity 9}, to obtain non-commutativity for closed string theory we need to examine T-dual theories of strings that propagate in coordinate dependent background fields. Usually this non-commutativity is obtained by examining the starting theory with coordinate dependent Kalb-Ramond field \cite{Lust 1,Adroit Larfos Lust Patalong,Androit Hohm Larfors Lust Patalong 1,Androit Hohm Larfors Lust Patalong 2,Blumenhagen Deser Plauschinn Rennecke Schmid 1,Blumenhagen Deser Plauschinn Rennecke Schmid 2,Blumenhagen Deser Plauschinn Rennecke Schmid 3,Lust 2,Blumenhagen Deser Lust Plauschinn Rennecke,Condeescu Florakis Lust,Shelton Taylor Wecht,Dabholkar Hull, Ilija 1, Ilija 2}. 

In this paper we examine a theory that has a coordinate dependent space-time metric tensor. This dependence will be linear and infinitesimal in coordinate $x^\mu$. The rest of the fields, Kalb-Ramond and dilaton field, will be set to zero. By applying Buscher T-duality procedure to this configuration of fields we will find T-dual theory. Since the metric tensor is dependent on coordinates and because this is a  symmetric tensor, we will have an emergence of $N(\xi)$ functions in transformation laws. These functions are generalization of standard $\beta_\mu$ functions that appear when we consider T-duality of theories that have coordinate dependent antisymmetric fields. Utilizing transformation laws that connect these two theories and enforcing standard Poisson bracket structure on starting theory we will examine Poisson bracket structure of dual theory. 

At the end we will give concluding remarks and in appendices we present definitions of $N(\xi)$ functions and some of their properties.

\section{Choice of background fields and action for closed bosonic string}
\setcounter{equation}{0}

Our starting point is the action for a closed bosonic string which propagates in presence of space-time metric $G(x)_{\mu\nu}$, Kalb-Ramond field $B(x)_{\mu\nu}$ and dilaton field $\Phi(x)$, where action is given as

\begin{equation}\label{action 1}
	S = \kappa \int_{\Sigma} d^2\xi \sqrt{-g} \left\{  \left[  \frac{1}{2} g^{m n} G_{\mu\nu}(x) + \frac{\epsilon^{mn}}{\sqrt{-g}} B_{\mu\nu} (x) \right]  \partial_m x^\mu \partial_n x^\nu  + \Phi(x) R^{(2)}    \right\},
\end{equation}

Here, $\Sigma$ is the world-sheet surface parameterized by $\xi^m = (\tau,\sigma)$ $[(m = 0,1), \sigma \in (0,\pi)]$, while the D-dimensional space-time is spanned by the coordinates $x^\mu \  (\mu = 0,1,2,...,D-1)$. Intrinsic world-sheet metric is labeled with $g_{mn}$ and its corresponding scalar curvature is given as $R^{(2)}$. 

Background fields $G(x)_{\mu\nu}$, $B(x)_{\mu\nu}$ and $\Phi(x)$ are not independent. They must satisfy certain equations of motion \cite{Callen Jr Martinec Perry Friedan}. These equations are a consequence of maintaining conformal symmetry at quantum level. They are given as

\begin{alignat}{2} \label{Pozadinska polja bozonska struna jed1}
	&\beta^G_{\mu\nu} &&\equiv R_{\mu\nu} - \frac{1}{4} B_{\mu\rho\sigma} B_{\nu}^{\rho\sigma} + 2 D_\mu a_\nu = 0,\\ \label{Pozadinska polja bozonska struna jed2}
	&\beta^B_{\mu\nu} &&\equiv D_\rho B^\rho_{\mu\nu} - 2 a_\rho B^\rho_{\mu\nu} = 0,\\ \label{Pozadinska polja bozonska struna jed3}
	&\beta^\Phi_{\ \ }  &&\equiv 2\pi\kappa \frac{D-26}{6} - R - \frac{1}{24} B_{\mu\rho\sigma} B^{\mu\rho\sigma} - D_\mu a^\mu + 4a^2 =c,
\end{alignat}
where we had that  $c$ is an arbitrary constant and function $\beta^\Phi$ could also be set to a constant because of the relation
\begin{equation}
	D^\nu \beta^G_{\nu\mu} + \partial_\mu \beta^\Phi = 0.
\end{equation}
Further, we also had that $R_{\mu\nu}$ and $D_\mu$ are Ricci tensors and covariant derivatives with respect to the space-time metric $G_{\mu\nu}$, 
\begin{align}
	R_{\mu\nu} &= \partial_\rho \Gamma^\rho_{\mu\nu} - \partial_\nu \Gamma^\rho_{\mu\rho} + \Gamma^\rho_{\delta\rho}\Gamma^\delta_{\mu\nu} - \Gamma^\rho_{\delta\nu}\Gamma^\delta_{\mu\rho},\\
	\Gamma^\rho_{\mu\nu} &= \frac{1}{2} G^{\rho\delta} \left( \partial_\mu G_{\delta\nu} + \partial_\nu G_{\mu\delta} - \partial_\delta G_{\mu\nu}  \right),\\
	D_\mu X^\nu &= \partial_\mu X^\nu + \Gamma^\nu_{\mu\rho} X^\rho, \quad D_\mu X_\nu = \partial_\mu X_\nu - \Gamma^\rho_{\mu\nu} X_\rho
\end{align}
where $X^\mu$ is an arbitrary tensor, while 
\begin{equation}
	B_{\mu\nu\rho} = \partial_\mu B_{\nu\rho} + \partial_\nu B_{\rho \mu} + \partial_\rho B_{\mu\nu}, \quad a_\mu = \partial_\mu \Phi,
\end{equation}	
 are field strength for Kalb-Ramond field $B_{\mu\nu}$ and dilaton gradient, respectively. Simplest possible solution to these equations is one where all background fields are set to constants. Since this solution does not possess any interesting properties, the next simplest one is where one of the fields has some infinitesimal and linear coordinate dependence. We will work with background fields where space-time metric has coordinate dependence and rest of the fields are set to zero.
Then, our background fields have the following form
 
\begin{equation}
	G(x)_{\mu\nu} = \eta_{\mu\nu} + h_{\mu\nu\rho} x^\rho, \quad B(x)_{\mu\nu} = 0,\quad \Phi(x) = 0,
\end{equation}
where $\eta_{\mu\nu}$ is flat space metric, $h_{\mu\nu\rho}$ is infinitesimal constant that is symmetric under exchange of all three indices.

Inserting this configuration of fields into the equations of motion for background fields (\ref{Pozadinska polja bozonska struna jed1})-(\ref{Pozadinska polja bozonska struna jed3}), we obtain 

\begin{equation}
	2\pi\kappa \frac{D-26}{6} = c.
\end{equation}

This equation allows us to specify in what number of dimensions we want to work in. Since we only wish to explore the main properties of this field configuration and in order not to complicate expressions needlessly we will set $D=3$. With this in mind we then have that

\begin{equation}
	G(x)_{\mu\nu} = \eta_{\mu\nu} + h_{\mu\nu\rho} x^\rho \quad \eta_{\mu\nu}  =  \begin{pmatrix}
		1 & 0 & 0\\
		0 & 1 & 0 \\
		0 & 0 & 1
	\end{pmatrix}.
\end{equation}

With this field configuration our action (\ref{action 1}) takes the following form

\begin{equation} \label{work act}
	S=\frac{k}{2}\int_{\Sigma}d^2\xi G(x)_{\mu\nu} \partial_+ x^\mu \partial_- x^\nu,
\end{equation}
where we have utilized world-sheet light-cone coordinates, more details in Appendix \ref{appendix A}. T-duality of a dilaton field is performed within quantum formalism, because of this we have neglected the term in which this field appears.

\section{Obtaining T-dual theory}
\setcounter{equation}{0}

Having defined our starting action, we are now in the position to start the process of T-dualization. In this section, we will focus on obtaining T-dual theory as well as T-dual transformation laws. We will utilize an extension of Buscher procedure that works on cases where we have coordinate dependent background fields and we do not posses translational symmetry.

\subsection{T-dualization}

Since action (\ref{work act}) is not invariant to translations, we begin by implementing following substitutions

\begin{alignat}{2}
	&\partial_\pm x^\mu     &&\rightarrow       v_\pm^\mu, \\ 
	&x^\rho                  &&\rightarrow       \Delta V^\rho    =     \int_{P} d { \xi^\prime }^m v_m^\rho (\xi^\prime)\label{V expresion 5},\\ 
	&S                    &&\rightarrow     S  +  \frac{\kappa}{2}    \int_{\Sigma}       d^2       \xi          \left[ v_+^\mu \partial_- y_\mu   -  v_-^\mu   \partial_+ y_\mu
	\right],
\end{alignat}
which produce following auxiliary action suitable for T-dualization

\begin{equation}
		S_{aux}=\frac{k}{2}\int_{\Sigma}d^2\xi \left[G(x)_{\mu\nu} v_+^\mu c_-^\nu + v_+^\mu \partial_- y_\mu - v_-^\mu \partial_+ y_\mu \right].
\end{equation}

We should note that path $P$, that appears in an integral for expression for $\Delta V^\rho$, starts from the point $\xi_0$ and ends in point $\xi$. This is the ingredient that makes auxiliary action non-local.

Now we will focus on finding the equations of motion for this action.  

Finding equations of motion for Lagrange multipliers
\begin{equation} \label{x form v 5}
	\partial_- v_+^\mu - \partial_+ v_-^\mu = 0, \quad v_\pm^\mu = \partial_\pm x^\mu,
\end{equation}
and inserting them into (\ref{V expresion 5}) we have
\begin{equation} \label{X from V 5}
	\Delta V^\rho = \int_{P} d { \xi^\prime }^m \partial_m x^\rho (\xi^\prime) = x^\rho (\xi) - x^\rho (\xi_0) = \Delta x^\rho.
\end{equation}
When translational symmetry is absent, in order to extract starting action from auxiliary one, we impose $x^\rho (\xi_0) =0$ as a constraint. Taking all this into account, we get the starting action (\ref{work act}).

Equations of motion for gauge fields are given as

\begin{align} \label{eq1}
	\partial_+ y_\mu(\kappa)  &= G(x)_{\mu\nu} x^\nu(\kappa)  + \int_{\Sigma} d^2 \xi v_+^\rho v_-^\gamma h_{\rho\gamma\nu} N(\kappa^-),\\ \label{eq2}
	\partial_- y_\mu(\kappa) &= - G(x)_{\mu\nu} v_-^\nu (\kappa) - \int_{\Sigma} d^2 \xi v_+^\rho v_-^\gamma h_{\rho\gamma\nu} N(\kappa^+).
\end{align}

 Functions $N(\kappa^\pm)$ are obtained from variation of term containing $\Delta V^\mu$. They are given as

\begin{align}
	N({\kappa^+}) =& 
	\delta      \Big(      { \xi^\prime }^- \big( ( { { \xi^\prime }^+ })^{-1}
	(\kappa^+) \big)   - \kappa^-
	\Big)     
	\left[    \bar{H}( \xi^+   - \kappa^+  )  -  H(\xi_0^+    -   \kappa^+)    \right], \\
	N(\kappa^-) 
	=&  
	\delta      \Big(      { \xi^\prime }^+ \big( ( { { \xi^\prime }^- })^{-1}
	(\kappa^-) \big)   - \kappa^+
	\Big)     
	\left[    \bar{H}( \xi^-   - \kappa^-  )  -  H(\xi_0^-    -   \kappa^-)    \right],
\end{align} 
more details on functions $N(\kappa^\pm)$, Dirac delta and step function are given in Appendix \ref{appendix B}.

Utilizing the fact that $h_{\mu\nu\rho}$ is infinitesimal, that is neglecting all terms that are of order $O(h^2)$, we can invert equations of motion (\ref{eq1}) and (\ref{eq2}), obtaining

\begin{align}
	v_-^\mu (\kappa) &= - G(x)^{\mu\nu} \left( \partial_- y_\nu (\kappa) - \int_{\Sigma} d^2 \xi \partial_+ y^\rho \partial_- y^\gamma h_{\rho\gamma\nu} N(\kappa^+)  \right), \\
	v_+^\mu (\kappa)&= G(x)^{\mu\nu} \left(  \partial_+ y_\nu (\kappa)  + \int_{\Sigma} d^2 \xi \partial_+ y^\rho \partial_- y^\gamma h_{\rho\gamma\nu} N(\kappa^-) \right),
\end{align}
where $y^\mu = \eta^{\mu\nu} y_{\nu}$.

Inserting these equations of motion into auxiliary action we obtain T-dual theory

\begin{equation}
	S_{T-dual} = \frac{k}{2} \int_{\Sigma} d^2\xi G^{\mu\nu} (\Delta V)^{\mu\nu} \partial_+ y_\mu \partial_- y_\nu.
\end{equation}

From here we can read T-dual space-time metric tensor

\begin{equation}
	{}^\star G(\Delta V)^{\mu\nu} =  G(\Delta V)^{\mu\nu} = \eta^{\mu\nu} - h^{\mu\nu}_\rho \Delta V^\rho.
\end{equation}

Comparing background fields of T-dual theory and starting theory we note that space-time metric tensor of dual theory is only the inverted metric tensor of a starting theory where coordinates have been replaced by non-local term $\Delta V$.

\subsection{Examination of geometric structure of T-dual theory}

In this chapter we focus on the examination of geometric properties of T-dual theory. This is accomplished by utilizing transformation laws that connect two theories and by imposing standard Poisson bracket structure on starting theory. Since original theory is geometric one, its Poisson brackets have following form
\begin{equation}
	\{ x^\mu (\sigma), \pi_\nu (\bar{\sigma})  \} = \delta^\mu_\nu \delta(\sigma - \bar{\sigma}), \quad \{ x^\mu(\sigma) , x^\nu(\bar{\sigma}) \} = 0, \quad \{ \pi_\mu(\sigma), \pi_\nu(\bar{\sigma})  \} = 0.
\end{equation}

In order to find Poisson brackets of dual theory we need to eliminate world-sheet proper time $\tau$ from T-dual transformation laws. One way do to this is to combine light-cone partial derivatives in a way that we end with a partial derivative with respect to $\sigma$ world-sheet parameter. We can utilize also momenta of starting theory to eliminate term $\partial_\tau x^\mu$. When we implement this we are still left with terms that contain $\partial_\tau$ which are tied to term containing $N(\kappa^\pm)$ functions. To eliminate these terms we need to utilze equations of motion of starting theory. Implementing all this we obtain following expression

\begin{equation}
	\partial_\sigma y_\mu = \frac{\pi_\mu}{k} - \frac{1}{2k} x^\rho \pi_\nu h^\nu_{\rho\mu},
\end{equation}
where 
\begin{equation}
	\pi_\mu = k G(x)_{\mu\nu} \partial_\tau x^\nu.
\end{equation}

To find Poisson brackets between T-dual coordinates we first need to find Poisson brackets between $\sigma$ partial derivatives of T-dual coordinates and then integrate the result with respect to $\sigma$ parameter. Doing this we obtain

\begin{equation}
	\{  \partial_\sigma y_\mu(\sigma) , \partial_{\bar{\sigma}} y_\nu (\bar{\sigma})  \} = 0, \quad \{ y_\mu (\sigma) , y_\nu (\bar{\sigma})   \} =C_{1 \mu} \bar{\sigma} + C_{2\nu} \sigma,
\end{equation}
where $C_{1\mu}$ and $C_{2\nu}$ are arbitrary constants, that can be set to zero.
Comparing this result to one obtained in Ref \cite{Nikolic Obric,Integration of Poisson brackets 2}, we notice that working with coordinate dependent space-time metric tensor instead of Kalb-Ramond tensor produces drastically different result.
T-dual theory, despite the fact that starting theory had coordinate dependent background fields, has, if we chose constants of integration to be zero, standard commutative geometric structure.

\section{Conclusion}
\setcounter{equation}{0}

In this article we examined T-dualization of bosonic string theory in presence of coordinate dependent space-time metric tensor. Where coordinate dependence was infinitesimal and linear, depending on parameter $h_{\mu\nu\rho} $ which was symmetric under exchange of indices. Rest of the fields, Kalb-Ramond and dilaton field were set to zero. This choice of the fields resembles one given in Ref \cite{Nikolic Obric,Integration of Poisson brackets 2} where roles of Kalb-Ramond and space-time metric tensor are switched. Because of this analogy we named this field configuration a "weakly curved space-time". Unlike the field configuration given in Ref \cite{Nikolic Obric,Integration of Poisson brackets 2}, this field configuration is not invariant under translational symmetry of coordinates $x^\mu$.

Since starting action possessed coordinate dependent background fields and because it lacked translational isometry, we needed to utilize generalization of Buscher procedure that was suitable for such cases. This procedure entailed that we replace starting action with one where partial derivatives have been replaced with covariant ones, coordinates have been replaced with invariant ones, excess degrees of freedom have been eliminated with Lagrange multipliers and where gauge freedom has been fixed. This auxiliary action is by its very construction non-local and it is the one on which we perform T-dualization procedure. To ensure that this auxiliary action is suitable replacement for our starting action we need to be able to obtain starting action from auxiliary action. This is done by finding equations of motion for Lagrange multipliers, solving them and inserting their solutions into auxiliary action. Choosing suitable starting point for the path that appears in invariant coordinate we are able to salvage starting theory. Finding equations of motion for gauge fields we obtain T-dual transformation laws. Inserting T-dual transformation laws into the auxiliary action we obtain T-dual action. Comparing starting and T-dual background fields we notice that Kalb-Ramond and dilaton fields have remained zero after dualization while T-dual space-time metric is inverse of space-time metric of original theory. In addition to this we now have that the parameter that appears in background fields is no longer coordinate $x^\mu$ but $\Delta V^\rho$.

Utilizing T-dual transformation laws we were able to deduce a transformation law for $\sigma$ partial derivative of T-dual coordinate $y_\mu$. By enforcing standard Poisson brackets onto starting theory and utilizing this transformation law we were able to find Poisson bracket between $\sigma$ partial derivatives of T-dual coordinate. It turns out that, unlike for the case where Kalb-Ramond field is coordinate dependent, this expression is zero. Integrating this expression twice we obtained Poisson bracket for dual theory. Poisson brackets of dual theory are dependent on arbitrary constant of integration which can be set to zero. By seting them to zero we obtained that dual theory has standard geometric structure. This result is in stark contrast with one obtained in Ref \cite{Nikolic Obric,Integration of Poisson brackets 2} and it hints at the possibility that in order to obtain non-commutative relations, one needs fields other than space-time metric to be coordinate dependent.

\appendix
\setcounter{equation}{0}

\section{Light-cone coordinates}
\label{appendix A}
Throughout this article we have utilized light-cone coordinates. In this appendix we will present some of the basic properties of these coordinates, defining them, defining their partial derivatives and expressing metric tensor in their basis. We begin by defining light-cone coordinates as
\begin{equation}
	\xi^\pm = \frac{1}{2} (\tau \pm \sigma).
\end{equation}
This definition naturally lends itself to introduction of corresponding partial derivatives
\begin{equation}
	\partial_\pm \equiv \frac{\partial}{\partial\xi^\pm} = \partial_\tau \pm \partial_\sigma.
\end{equation}
Having defined light-cone coordinates we can take a look at the metric tensor. Here we wrote down metric tensor in $(\tau,\sigma)$ and light-cone basis. respectively
\begin{equation}
	\eta = \begin{pmatrix}
		1 & 0 \\
		0 & -1
	\end{pmatrix}, \quad
	\eta_{lc} = \begin{pmatrix}
		0 & \frac{1}{2}\\
		\frac{1}{2} & 0
	\end{pmatrix}.
\end{equation}
Here, $lc$ denotes light-cone basis.

\section{Derivation of $N(\kappa^\pm)$ functions}
\label{appendix B}
\setcounter{equation}{0}
In this article we had to find variation of term that contained $Delta V$. This gave rise to $N(\kappa^\pm)$ functions. Here we will give systematic derivation of these functions, we will also present some of their properties. We obtain these functions in thew following manner

\begin{align}
	\begin{gathered}
		 v_+^\alpha v_-^\beta\frac{ \delta G(\Delta V)_{\alpha\beta}  }{ \delta v_+^\mu ( \kappa)} =
		v_+^\alpha v_-^\beta h_{\alpha\beta\rho} \frac{\delta \Delta V^\rho}{\delta v_+^\mu ( \kappa)} = 
		v_+^\alpha v_-^\beta h_{\alpha\beta\rho} \int_{P} d \xi^{\prime m} \frac{ \delta v_m^\rho (\xi^\prime)}{\delta v_+^\mu ( \kappa)}  = \\
		\begin{aligned}
			&=v_+^\alpha v_-^\beta h_{\alpha\beta\mu} 
			\int_{P} d { \xi^\prime }^+ \delta( { \xi^\prime }^+   - \kappa^+ )       \delta(   { \xi^\prime }^-   - \kappa^-  ) \\
			& =v_+^\alpha v_-^\beta h_{\alpha\beta\mu}   
			\int_{t_i}^{t_f} dt \frac{d{ \xi^\prime }^+}{dt} \delta( { \xi^\prime (t) }^+   - \kappa^+ )       \delta(   { \xi^\prime }^- (t)   - \kappa^-  ) \\
			& = v_+^\alpha v_-^\beta h_{\alpha\beta\mu}  
			\int_{{\xi_0 }^+}^{{\xi }^+} du  \delta( u   - \kappa^+ )       \delta      \Big(      { \xi^\prime }^- \big( ( { { \xi^\prime }^+ })^{-1}
			(u) \big)   - \kappa^-
			\Big)    \\
			&= v_+^\alpha v_-^\beta h_{\alpha\beta\mu} 
			\delta      \Big(      { \xi^\prime }^- \big( ( { { \xi^\prime }^+ })^{-1}
			(\kappa^+) \big)   - \kappa^-
			\Big)     
			\left[    \bar{H}( \xi^+   - \kappa^+  )  -  \bar{H}(\xi_0^+    -   \kappa^+)    \right]\\
			&=v_+^\alpha v_-^\beta h_{\alpha\beta\mu}  N(\kappa^+).
		\end{aligned}
	\end{gathered}
\end{align}

In the third line we have parametrised the path integral with parameter $t$ where ${\xi^\prime }^+(t_i) ={\xi_0 }^+$ and ${\xi^\prime }^+(t_f) ={\xi }^+$. In the fourth line we have introduced the substitution $u={\xi^\prime }^+(t)$. The fifth line is obtained by using following integration rule for Dirac delta function

\begin{equation}
	\int_{\sigma_0}^{\sigma} d\eta f(\eta) \delta(\eta - \bar{\eta}) = f(\bar{\eta}) \left[ \bar{H}(\sigma - \bar{\eta})  - \bar{H}(\sigma_0 - \bar{\eta})   \right].
\end{equation}
Here, $\bar{H}(x)$ is a step function defined as
\begin{equation}\label{eq:fdelt}
	\bar{H}(x)=\int_0^x d\eta\delta(\eta)=\frac{1}{2\pi}\left[x+2\sum_{n\ge 1}\frac{1}{n}\sin(nx)\right]=\left\{\begin{array}{ll}
		0 & \textrm{if $x=0$}\\
		1/2 & \textrm{if $0<x<2\pi$}\, ,\\
		1 & \textrm{if $x=2\pi$} \end{array}\right .
\end{equation}

Procedure for obtaining the $N(\kappa^-)$ term is analogous.

Now we will list some of the properties of these functions

Functions $N(\kappa^\pm)$ can be combined in the following way to obtain functions that are dependent on $\tau$ and $\sigma$ coordinates

\begin{gather}
	N(\kappa^+) + N(\kappa^-) = N(\kappa^0),\\
	N(\kappa^+) - N(\kappa^-) = N(\kappa^1),
\end{gather}
where $\kappa^0$ and $\kappa^1$ represent $\tau$ and $\sigma$ coordinates respectively.

Acting with partial derivatives on $N(\kappa^\pm)$ functions and then integrating over world-sheet we obtain following set of equations
\begin{gather}
	\int_{\Sigma} d^2 \xi \partial_+ N(\kappa^+)  = 1, \qquad
	\int_{\Sigma} d^2 \xi \partial_- N(\kappa^+)  = 0,\\
	\int_{\Sigma} d^2 \xi \partial_- N(\kappa^-)  = 1, \qquad
	\int_{\Sigma} d^2 \xi \partial_+ N(\kappa^-)  = 0.
\end{gather}

Same applies for $N(\kappa^0)$ and $N(\kappa^1)$ functions.

\end{document}